\documentclass{article}
\usepackage{amsfonts}
\usepackage{amsmath}
\usepackage{cite}

\input{tcilatex}

\begin{document}

\title{Auto-correlation Function Study of Scattered Light Intensity}
\author{Yong Sun{\thanks{%
Email: ysun2002h@yahoo.com.cn}}}
\maketitle

\begin{abstract}
In this work, the particle size distribution measured using the dynamic
light scattering (DLS) technique is compared with that obtained from the
static light scattering (SLS) technique or provided by the supplier measured
using the Transmission Electron Microscopy (TEM) technique for dilute Poly($%
N $-isopropylacrylamide) microgel and standard polystyrene latex samples in
dispersion respectively. The results show that the narrow particle size
distribution that can be measured accurately using the SLS technique is not
suited to the determination by the DLS technique and the particle size
distribution obtained from the DLS technique is different from the value
provided by the supplier. With the assistance of the simulated data of the
normalized time auto-correlation function of the scattered light intensity $%
g^{\left( 2\right) }\left( \tau \right) $, the effects of the particle size
distribution on the nonexponentiality of $g^{\left( 2\right) }\left( \tau
\right) $ measured at a scattering angle of 30$^{\text{o}}$ are
investigated. The analysis reveals that the influences of the particle size
distribution are small on the nonexponentiality of $g^{\left( 2\right)
}\left( \tau \right) $ and very large on the initial slope of the logarithm
of $g^{\left( 2\right) }\left( \tau \right) $. The values of the apparent
hydrodynamic radius are also largely influenced by the particle size
distribution and the difference between the distributions of the apparent
hydrodynamic radius and hydrodynamic radius of particles is determined by
the method of cumulants.
\end{abstract}

\section{INTRODUCTION}

For colloidal dispersion systems, light scattering is a widely used
technique to measure the sizes of particles. One of the main applications of
the dynamic light scattering (DLS) technique is to measure the sizes of
spherical particles in liquid suspension. The standard method of cumulants%
\cite{re1,re2,re3,re4} has been used to measure the hydrodynamic radius, or
more strictly apparent hydrodynamic radius $R_{h,app}$ \cite{re5} of
particles from the normalized time auto-correlation function of the
scattered light intensity $g^{\left( 2\right) }\left( \tau \right) $ with
the assistance of the Einstein-Stokes relation, where $\tau $ is the delay
time. However this method is insensitive to a small poly-dispersity and not
suited to the accurate determination for the poly-dispersities (standard
deviation/mean size) less than about 10\%. For large particles, DLS
technique, where it loses the accuracy of size measurements, is endeavored
to use at different scattering angles in order to obtain the effective
diffusion coefficient\cite{re6} or the apparent hydrodynamic radius\cite{re5}
to detect small poly-dispersities.

In the previous work\cite{re7,re8}, I have discussed that a new radius,
static radius, can be obtained from the static light scattering (SLS)
technique, the particle size distribution less than 10\% can be measured
accurately and the three different particle sizes, static, hydrodynamic and
apparent hydrodynamic radii, can be obtained using the light scattering
technique. In this work, the particle size distribution measured using the
DLS technique is compared with that obtained from the SLS technique or
provided by the supplier measured using the Transmission Electron Microscopy
(TEM) technique for dilute Poly($N$-isopropylacrylamide) (PNIPAM) microgel
and standard polystyrene latex samples in dispersion respectively. The
results show that the narrow particle size distribution that can be measured
accurately using the SLS technique is not suited to the determination by the
DLS techniques and the particle size distribution obtained using the DLS
technique is different from the value provided by the supplier. Since the
method of cumulants measures the particle size distribution from the
deviation between the exponentiality and $g^{\left( 2\right) }\left( \tau
\right) $ measured at a single scattering angle, the simulated data were
thus used to explore the effects of the particle size distribution on the
difference between them at a scattering angle of 30$^{\text{o}}$. The
analysis reveals that the effects of the particle size distribution are
small on the nonexponentiality of $g^{\left( 2\right) }\left( \tau \right) $
and very large on the initial slope of the logarithm of $g^{\left( 2\right)
}\left( \tau \right) $. The values of the apparent hydrodynamic radius are
also largely influenced by the particle size distribution and the difference
between the distributions of the apparent hydrodynamic radius and
hydrodynamic radius of particles is determined by the method of cumulants.

\section{THEORY}

For dilute poly-disperse homogeneous spherical particles in dispersion where
the Rayleigh-Gans-Debye (RGD) approximation is valid, the normalized time
auto-correlation function of the electric field of the scattered light $%
g^{\left( 1\right) }\left( \tau \right) $ is given by

\begin{equation}
g^{\left( 1\right) }\left( \tau \right) =\frac{\int_{0}^{\infty
}R_{s}^{6}P\left( q,R_{s}\right) G\left( R_{s}\right) \exp \left(
-q^{2}D\tau \right) dR_{s}}{\int_{0}^{\infty }R_{s}^{6}P\left(
q,R_{s}\right) G\left( R_{s}\right) dR_{s}},  \label{Grhrs}
\end{equation}
where $R_{s}$ is the static radius, $D$ is the diffusion coefficient, $\ q=%
\frac{4\pi }{\lambda }n_{s}\sin \frac{\theta }{2}$ is the scattering vector, 
$\lambda $\ is the wavelength of the incident light in vacuo, $n_{s}$ is the
solvent refractive index, $\theta $ is the scattering angle, $G\left(
R_{s}\right) $ is the number distribution of particle sizes and the form
factor $P\left( q,R_{s}\right) $ is

\begin{equation}
P\left( q,R_{s}\right) =\frac{9}{q^{6}R_{s}^{6}}\left( \sin \left(
qR_{s}\right) -qR_{s}\cos \left( qR_{s}\right) \right) ^{2}.  \label{factor}
\end{equation}

\noindent From the Einstein-Stokes relation

\begin{equation}
D=\frac{k_{B}T}{6\pi \eta _{0}R_{h}},
\end{equation}
where $\eta _{0}$, $k_{B}$ and $T$ are the viscosity of the solvent,
Boltzmann's constant and absolute temperature respectively, the hydrodynamic
radius $R_{h}$ can be obtained.

Traditionally the cumulants is a standard method to measure the particle
size distribution from the DLS data $g^{\left( 2\right) }\left( \tau \right)$%
. In this work, the following equation was used to analyze the DLS data to
the second moment

\begin{equation}
g^{\left( 2\right) }\left( \tau \right) =1+\beta \exp \left( -2\left\langle
\Gamma \right\rangle \tau \right) \left( 1+\mu _{2}\tau ^{2}\right) ,
\label{cum}
\end{equation}%
where $\left\langle \Gamma \right\rangle =q^{2}D_{e}\left( q\right) $ is the
average decay rate, $D_{e}\left( q\right) $ is the effective diffusion
coefficient, $\mu _{2}$ is the second moment and $\beta $ is a constant that
depends on the experimental geometry for a given experimental measurement.
The apparent hydrodynamic radius $R_{h,app}$ can be obtained from $%
D_{e}\left( q\right) $

\begin{equation}
R_{h,app}=\frac{k_{B}T}{6\pi \eta _{0}D_e} .  \label{Rhapp}
\end{equation}

\noindent The relative width of the apparent hydrodynamic radius
distribution is\cite{re9} 
\begin{equation}
\frac{Width}{R_{h,app}}=\frac{\sqrt{\mu _2}}{\left\langle \Gamma
\right\rangle}.  \label{Rhwidth}
\end{equation}

\noindent If the first cumulant is used, the value of the apparent
hydrodynamic radius $R_{h,app}$ at a given scattering angle and a given
delay time $\tau$ can be calculated directly using the static particle size
information and the relationship between the static and hydrodynamic radii.
If the DLS data during the delay time range $\tau _1$ and $\tau _2$ are
chosen to obtain $R_{h,app}$ at a given scattering angle, the average value
of apparent hydrodynamic radius can be calculated using the following
equation

\begin{equation}
R_{h,app}\left( e^{-\frac{q^{2}k_{B}T\tau _{1}}{6\pi \eta _{0}R_{h,app}}%
}-e^{-\frac{q^{2}k_{B}T\tau _{2}}{6\pi \eta _{0}R_{h,app}}}\right) =\frac{%
\int_{0}^{\infty }R_{h}R_{s}^{6}P\left( q,R_{s}\right) G\left( R_{s}\right)
\left( e^{-\frac{q^{2}k_{B}T\tau _{1}}{6\pi \eta _{0}R_{h}}}-e^{-\frac{%
q^{2}k_{B}T\tau _{2}}{6\pi \eta _{0}R_{h}}}\right) dR_{s}}{\int_{0}^{\infty
}R_{s}^{6}P\left( q,R_{s}\right) G\left( R_{s}\right) R_{s}}.
\label{Rhallcal}
\end{equation}

\section{EXPERIMENT}

The SLS and DLS data were measured using the instrument built by ALV-Laser
Vertriebsgesellschaft m.b.H (Langen, Germany). It utilizes an ALV-5000
Multiple Tau Digital Correlator and a JDS Uniphase 1145P He-Ne laser to
provide a 23 mW vertically polarized laser at wavelength of 632.8 nm.

In this work, two kinds of samples were used. One is PNIPAM submicron
spheres and the other is standard polystyrene latex spheres. The samples
used in this work have been detailed before\cite{re7}. The four PNIPAM
microgel samples PNIPAM-0, PNIPAM-1, PNIPAM-2 and PNIPAM-5 were named
according to the molar ratios $n_{B}/n_{N}$ of cross-linker $N,N^{\prime }$%
-methylenebisacrylamide over $N$-isopropylacrylamide. The sulfate
polystyrene latex with a normalized mean radius of 33.5 nm and
surfactant-free sulfate polystyrene latex of 55 nm were named Latex-1 and
Latex-2 respectively.

\section{DATA ANALYSIS}

In this section, the particle size information obtained using the method of
cumulants from the DLS data is compared with the commercial values for the
standard polystyrene latex samples and the values obtained using the SLS
technique for the PNIPAM submicron spheres.

\subsection{Standard polystyrene latex samples}

When Eq. \ref{cum} was used to fit the data of Latex-1 measured at a
temperature of 298.45 K and a scattering angle of 30$^{\text{o}}$ under the
conditions of $\mu _{2}=0$ and $\mu _{2}\neq 0$ respectively, it was found
that the results of $\left\langle \Gamma \right\rangle $ and $\mu _{2}$
depend on the delay time range being fit, as shown in Table \ref{Latex-130}.
If a small delay time range is chosen, the parameters are not
well-determined. As the delay time range is increased, the uncertainties in
parameters decrease and $\left\langle \Gamma \right\rangle $ and $\mu _{2}$
stabilize. From Eq. \ref{Rhwidth}, the relative width of the apparent
hydrodynamic radius distribution is about 0.26.

\begin{table}[th]
\begin{center}
\begin{tabular}{|c|c|c|c|c|c|}
\hline
Delay time (s) & $\left\langle\Gamma \right\rangle_{first}\left(
s^{-1}\right) $ & $\chi ^{2}$ & $\left\langle\Gamma\right\rangle
_{two}\left( s^{-1}\right) $ & $\mu _{2}\left( s^{-2}\right) $ & $\chi ^{2}$
\\ \hline
2$\times$10$^{-7}$ to 0.00102 & 315.6 $\pm$ 1.2 & 0.34 & 323.1 $\pm$ 3.4 & 
21000 $\pm$ 7000 & 0.27 \\ \hline
2$\times$10$^{-7}$ to 0.00154 & 313.5 $\pm$ 0.9 & 0.40 & 321.1 $\pm$ 2.4 & 
15000 $\pm$ 4000 & 0.26 \\ \hline
2$\times$10$^{-7}$ to 0.00364 & 311.1 $\pm$ 0.7 & 0.60 & 319.7 $\pm$ 1.8 & 
12000 $\pm$ 2000 & 0.26 \\ \hline
2$\times$10$^{-7}$ to 0.00364 & 310.0 $\pm$ 0.6 & 0.73 & 318.3 $\pm$ 1.4 & 
9000 $\pm$ 1000 & 0.27 \\ \hline
2$\times$10$^{-7}$ to 0.00568 & 309.4 $\pm$ 0.6 & 0.81 & 317.1 $\pm$ 1.2 & 
8000 $\pm$ 1000 & 0.28 \\ \hline
2$\times$10$^{-7}$ to 0.00896 & 309.3 $\pm$ 0.6 & 0.80 & 316.6 $\pm$ 1.2 & 
7000 $\pm$ 1000 & 0.29 \\ \hline
2$\times$10$^{-7}$ to 0.01306 & 309.3 $\pm$ 0.6 & 0.82 & 316.8 $\pm$ 1.2 & 
7000 $\pm$ 1000 & 0.31 \\ \hline
\end{tabular}%
\end{center}
\caption{The fit results obtained using Eq. \protect\ref{cum} with $\protect%
\mu _{2}=0$ and $\protect\mu _{2}\neq 0$ respectively for Latex-1 at
different delay time ranges, a scattering angle of 30$^{\text{o}}$ and a
temperature of 298.45 K.}
\label{Latex-130}
\end{table}

Fit values obtained using both procedures are listed in Table \ref%
{5Latex-130} for five independent DLS data measured at a scattering angle of
30$^{\text{o}}$ and during the delay time range from 2$\times $10$^{-7}$ to
0.00896 s. The fit results show that the average values of decay rate are
consistent and the values of $\mu _{2}$ have some difference. The relative
width of the apparent hydrodynamic radius is from 0.23 to 0.28. However,
from the particle size information provided by the supplier, the relative
width of the radius distribution is only 0.07. The width of distribution
obtained using the DLS technique at a scattering angle of 30$^{\text{o}}$ is
much larger than that obtained using TEM technique. The value of the
apparent hydrodynamic radius is 37.29$\pm $0.08 nm using the first cumulant
method and 36.5$\pm $0.2 nm using the first two cumulant method. The value
of the apparent hydrodynamic radius obtained from the DLS technique for
Latex-1 at a scattering angle of 30$^{\text{o}}$ is not equal to the mean
radius 33.5 nm provided by the supplier.

\begin{table}[th]
\begin{center}
\begin{tabular}{|c|c|c|c|c|c|}
\hline
& $\left\langle\Gamma\right\rangle _{first}\left( s^{-1}\right) $ & $\chi
^{2}$ & $\left\langle\Gamma\right\rangle _{two}\left( s^{-1}\right) $ & $\mu
_{2}\left( s^{-2}\right) $ & $\chi ^{2}$ \\ \hline
1 & 308.4$\pm $0.6 & 0.69 & 314.$\pm $1. & 5000$\pm $1000 & 0.42 \\ \hline
2 & 310.1$\pm $0.6 & 0.75 & 318.$\pm $1. & 7000$\pm $1000 & 0.18 \\ \hline
3 & 309.3$\pm $0.6 & 0.80 & 317.$\pm $1. & 7000$\pm $1000 & 0.29 \\ \hline
4 & 308.8$\pm $0.6 & 0.88 & 317.$\pm $1. & 8000$\pm $1000 & 0.25 \\ \hline
5 & 309.9$\pm $0.6 & 0.64 & 316.$\pm $1. & 6000$\pm $1000 & 0.27 \\ \hline
\end{tabular}%
\end{center}
\caption{The fit results obtained using Eq. \protect\ref{cum} with $\protect%
\mu _{2}=0$ and $\protect\mu _{2}\neq 0$ respectively for the five
independent DLS data of Latex-1 measured at a scattering angle of 30$^{\text{%
o}}$ and a temperature of 298.45 K.}
\label{5Latex-130}
\end{table}

The intensity-intensity correlation function $g^{\left( 2\right) }\left(
\tau \right) $ measured at a scattering angle 30$^{\text{o}}$ is shown in
Fig. 1. Figure 1a shows the data of $g^{\left( 2\right) }\left( \tau \right) 
$ and a fit of Eq. \ref{cum} to the data with $\mu _{2}=0$ during the delay
time range from 2$\times $10$^{-7}$ to 0.00896 s. The residuals $\left(
y_{i}-y_{fit}\right) /\sigma _{i}$ show systematic variations with the delay
time, where $y_{i}$, $y_{fit}$ and $\sigma _{i}$ are the data, the fit value
and the uncertainty in the data at a given delay time $\tau _{i}$,
respectively. Figure 1b shows the same data with a fit of Eq. \ref{cum} in
which $\mu _{2}\neq 0$. The residuals show systematic variations again.

When Eq. \ref{cum} was used to fit the DLS data of Latex-1 measured at a
scattering angle of 90$^{\text{o}}$ and the same temperature for $\mu _{2}=0$
and $\mu _{2}\neq 0$ respectively, it was found that the results of $%
\left\langle \Gamma \right\rangle $ and $\mu _{2}$ depend on the delay time
range being fit, as shown in Table \ref{Latex-190}. As the delay time range
is increased, the uncertainties in parameters decrease. However the values
of $\left\langle \Gamma \right\rangle $ still do not stabilize and the value
of $\mu _{2}$ not only has a strong dependence on the delay time range being
fit but also is negative. It's a contradiction with its definition. For the
five independent DLS data measured at a scattering angle of 90$^{\text{o}}$,
the values of $\mu _{2}$ also show a strong dependence on the DLS
measurements and the width of the apparent hydrodynamic radius cannot be
determined at this scattering angle.

\begin{table}[th]
\begin{center}
\begin{tabular}{|c|c|c|c|c|c|}
\hline
Delay time (s) & $\left\langle\Gamma\right\rangle_{first}\left(
s^{-1}\right) $ & $\chi ^{2}$ & $\left\langle\Gamma\right\rangle_{two}\left(
s^{-1}\right) $ & $\mu _{2}\left( s^{-2}\right) $ & $\chi ^{2}$ \\ \hline
2$\times$10$^{-7}$ to 1.92$\times$10$^{-4}$ & 2310 $\pm$ 14 & 0.46 & 2252 $%
\pm$ 47 & -705056 $\pm$ 533695 & 0.44 \\ \hline
2$\times$10$^{-7}$ to 3.072$\times$10$^{-4}$ & 2314 $\pm$ 10 & 0.45 & 2290 $%
\pm$ 30 & -210755 $\pm$ 241864 & 0.45 \\ \hline
2$\times$10$^{-7}$ to 4.608$\times$10$^{-4}$ & 2315 $\pm$ 9 & 0.43 & 2301 $%
\pm$ 22 & -91973 $\pm$ 135873 & 0.43 \\ \hline
2$\times$10$^{-7}$ to 7.168$\times$10$^{-4}$ & 2316 $\pm$ 8 & 0.42 & 2301 $%
\pm$ 18 & -87254 $\pm$ 90911 & 0.41 \\ \hline
2$\times$10$^{-7}$ to 0.00113 & 2317 $\pm$ 8 & 0.45 & 2293 $\pm$ 16 & 
-138033 $\pm$ 76731 & 0.42 \\ \hline
\end{tabular}%
\end{center}
\caption{The fit results obtained using Eq. \protect\ref{cum} with $\protect%
\mu _{2}=0$ and $\protect\mu _{2}\neq 0$ respectively for Latex-1 at
different delay time ranges, a scattering angle of 90$^{\text{o}}$ and a
temperature of 298.45 K.}
\label{Latex-190}
\end{table}

Figure 2a shows the normalized time auto-correlation function of the
scattered light intensity $g^{\left( 2\right) }\left( \tau \right) $ for
Latex-1 at a scattering angle of 90$^{\text{o}}$. Equation \ref{cum} was fit
to the data with $\mu _{2}=0$. The residuals vary randomly as the delay time
is changed. Figure 2b shows the same data with a fit of Eq. \ref{cum} in
which $\mu _{2}\neq 0$. Again, the residuals are also random. The value of
the apparent hydrodynamic radius is 37.4$\pm $0.1 nm using the first
cumulant method and 37.5$\pm $0.3 nm using the first two cumulant method.
The value of the apparent hydrodynamic radius obtained from the DLS
technique for Latex-1 at a scattering angle of 90$^{\text{o}}$ is still not
equal to the mean radius 33.5 nm provided by the supplier. For the other
polystyrene latex sample, the situation is the same: the width of the
apparent hydrodynamic radius cannot be determined using this DLS technique
and the value of the apparent hydrodynamic radius is larger than that
provided by the supplier.

\subsection{PNIPAM samples}

Equation \ref{cum} was also used to obtain decay rates for PNIPAM samples
with $\mu _{2}=0$ and $\mu _{2}\neq 0$ respectively. Figure 3a shows the
results of fitting Eq. \ref{cum} to the PNIPAM-1 data with $\mu _{2}=0$ over
the delay time range 2$\times $10$^{-7}$ to 0.06548 s measured at a
scattering angle of 30$^{\text{o}}$ and a temperature of 302.38K. The
residuals show systematic variations with the delay time. Figure 3b shows
the same data with a fit of Eq. \ref{cum} in which $\mu _{2}\neq 0$. The
residuals also show systematic variations with the delay time. Fit values
obtained using the both procedures for the five independent data sets are
listed in Table \ref{PNIPAM-129} at a scattering angle of 30$^{\text{o}}$
and a temperature of 302.38K. The results of $\mu _{2}$ show a strong
dependence on the DLS measurements and the width of the apparent
hydrodynamic radius cannot be determined. Using the SLS technique, the
relative width of the static radius can be measured accurately and is 0.085.
The value of the apparent hydrodynamic radius is 323.$\pm $2. nm using the
first cumulant method and 319.$\pm $2. nm using the first two cumulant
method. The values are much larger than the value 254.3$\pm $0.1 nm obtained
from the SLS technique\cite{re7}. For the other PNIPAM samples investigated,
the situation is the same: the distributions that can be measured accurately
using the SLS technique cannot be determined by the cumulants method and the
value of the hydrodynamic radius obtained using the cumulants is larger than
that obtained from the SLS technique.

\begin{table}[ht]
\begin{center}
\begin{tabular}{|c|c|c|c|c|c|}
\hline
& $\left\langle\Gamma \right\rangle_{first}\left( s^{-1}\right) $ & $\chi
^{2}$ & $\left\langle\Gamma \right\rangle_{two}\left( s^{-1}\right) $ & $\mu
_{2}\left( s^{-2}\right) $ & $\chi ^{2}$ \\ \hline
1 & 39.65 $\pm $ 0.07 & 0.27 & 39.8 $\pm $ 0.1 & 17 $\pm $ 14 & 0.26 \\ 
\hline
2 & 39.39 $\pm $ 0.07 & 0.67 & 40.1 $\pm $ 0.1 & 94 $\pm $ 15 & 0.35 \\ 
\hline
3 & 39.77 $\pm $ 0.07 & 0.46 & 40.0 $\pm $ 0.1 & 35 $\pm $ 14 & 0.41 \\ 
\hline
4 & 39.62 $\pm $ 0.07 & 0.36 & 39.8 $\pm $ 0.1 & 17 $\pm $ 14 & 0.35 \\ 
\hline
5 & 39.22 $\pm $ 0.07 & 1.09 & 40.2 $\pm $ 0.1 & 127$\pm $ 16 & 0.50 \\ 
\hline
\end{tabular}
\makeatletter
\par
\makeatother 
\end{center}
\caption{The fit results obtained using Eq. \protect\ref{cum} with $\protect%
\mu_2=0$ and $\protect\mu_2 \neq 0$ respectively for the five independent
DLS data of PNIPAM-1 at a scattering angle of 30$^\mathrm o$ and a
temperature of 302.38K.}
\label{PNIPAM-129}
\end{table}

\section{RESULTS AND DISCUSSION}

Because the expected values of the DLS data calculated based on the
commercial and static particle size information are consistent with the
experimental data\cite{re7,re8}, the DLS simulated data were used to explore
the effects of particle size distribution on the deviation between the
exponentiality and $g^{\left( 2\right) }\left( \tau \right) $ and the
initial slope of the logarithm of $g^{\left( 2\right) }\left( \tau \right) $
at a scattering angle of 30$^{\text{o}}$. The method that produces the DLS
simulated data has been detailed before\cite{re7}. In this work, the number
distribution of particle sizes is still chosen as a Gaussian distribution

\begin{equation}
G\left( R_{s};\left\langle R_{s}\right\rangle ,\sigma \right) =\frac{1}{
\sigma \sqrt{2\pi }}\exp \left( -\frac{1}{2}\left( \frac{R_{s}-\left\langle
R_{s}\right\rangle }{\sigma }\right) ^{2}\right) ,
\end{equation}
where $\left\langle R_{s}\right\rangle $ is the mean static radius and $%
\sigma $ is the standard deviation related to the mean static radius.

The simulated data were produced using the information: the mean static
radius $\left\langle R_{s}\right\rangle $, standard deviation $\sigma $,
temperature $T$, viscosity of the solvent $\eta _{0}$, scattering angle $%
\theta $, wavelength of laser light $\lambda $, refractive index of the
water $n_{s}$ and constant $a=R_{h}/R_{s}$ were set to 50 nm, 10 nm,
300.49K, 0.8479 mPa$\cdot $S, 30$^{\text{o}}$, 632.8 nm, 1.332 and 1.1,
respectively. When the data of $\left( g^{\left( 2\right) }\left( \tau
\right) -1\right) /\beta $ were obtained, the 1\% statistical noises were
added and the random errors were set 3\%. Five simulated data were produced
respectively. The fit results for one of the DLS simulated data at different
delay time ranges using Eq. \ref{cum} with $\mu _{2}=0$ and $\mu _{2}\neq 0$
respectively are listed in Table \ref{data5010}.

\begin{table}[th]
\begin{center}
\begin{tabular}{|c|c|c|c|c|c|}
\hline
Delay time range $\left( s\right) $ & $\left\langle\Gamma\right\rangle
_{first}\left( s^{-1}\right) $ & $\chi ^{2}$ & $\left\langle\Gamma\right%
\rangle _{two}\left( s^{-1}\right) $ & $\mu _{2}\left( s^{-2}\right) $ & $%
\chi ^{2}$ \\ \hline
2$\times$10$^{-7}$ to 0.00402 & 186.78 $\pm$ 0.09 & 1.09 & 188.9 $\pm$ 0.5 & 
1237 $\pm$ 265 & 0.98 \\ \hline
2$\times$10$^{-7}$ to 0.00602 & 186.67 $\pm$ 0.08 & 1.11 & 188.3 $\pm$ 0.3 & 
843 $\pm$ 158 & 0.97 \\ \hline
2$\times$10$^{-7}$ to 0.00802 & 186.60 $\pm$ 0.07 & 1.14 & 188.1 $\pm$ 0.2 & 
768 $\pm$ 116 & 0.93 \\ \hline
2$\times$10$^{-7}$ to 0.01002 & 186.18 $\pm$ 0.07 & 1.85 & 188.2 $\pm$ 0.2 & 
786 $\pm$ 56 & 0.92 \\ \hline
2$\times$10$^{-7}$ to 0.01202 & 185.67 $\pm$ 0.06 & 3.16 & 188.3 $\pm$ 0.1 & 
842 $\pm$ 39 & 0.92 \\ \hline
2$\times$10$^{-7}$ to 0.01402 & 184.42 $\pm$ 0.05 & 7.90 & 188.4 $\pm$ 0.1 & 
905 $\pm$ 24 & 0.92 \\ \hline
2$\times$10$^{-7}$ to 0.01602 & 183.44 $\pm$ 0.04 & 12.51 & 188.4 $\pm$ 0.1
& 904 $\pm$ 19 & 0.88 \\ \hline
\end{tabular}%
\end{center}
\caption{The fit results of simulated data produced based on the mean static
radius 50 nm and standard deviation 10 nm at different delay time ranges
using Eq. \protect\ref{cum} with $\protect\mu _{2}=0$ and $\protect\mu %
_{2}\neq 0$ respectively.}
\label{data5010}
\end{table}

The fit results at different delay time ranges shown in Table \ref{data5010}
are the same situation as the results of Latex-1 at a scattering angle of 30$%
^{\text{o}}$. When Eq. \ref{cum} was used to fit the data of the simulated
data produced based on the mean static radius 50 nm and standard deviation
10 nm, it was found that the results of $\left\langle \Gamma \right\rangle $
and $\mu _{2}$ depend on the delay time range being fit. If a small delay
time range is chosen, the parameters are not well-determined. As the delay
time range is increased, the uncertainties in parameters decrease and $%
\left\langle \Gamma \right\rangle $ and $\mu _{2}$ stabilize. The fit
results for $g^{\left( 2\right) }\left( \tau \right) $ are shown in Fig. 4.
For both the fit results, the residuals are random as the delay time is
changed. From Eq. \ref{Rhwidth}, the relative width of the apparent
hydrodynamic radius is about 0.16. It is different from the relative width
of this simulated data 0.2. The fit results for the five simulated data also
show that the values of mean decay constant $\left\langle \Gamma
\right\rangle $ are consistent and the results of $\mu _{2}$ have some
differences for the different noises and errors like the values of $\mu _{2}$
of Latex-1 at a scattering angle of 30$^{\text{o}}.$

The fit results for other particle size distributions also have been
analyzed. The simulated data were produced using the same temperature $T$,
viscosity of the solvent $\eta _{0}$, scattering angle $\theta $, wavelength
of laser light $\lambda $ and refractive index of the water $n_{s}$. The
constant $a$ for the mean static radius 50 nm and standard deviations 3 nm,
5 nm, 15 nm, 20 nm and 25 nm was chosen to 1.1 and for the mean static
radius 260 nm and standard deviations 16 nm, 26 nm, 52 nm, 78 nm, 104 nm and
130 nm was set to 1.2, respectively. The results fit Eq. \ref{cum} to one of
the simulated data produced using the mean static radius 50 nm and standard
deviation 3 nm at different delay time ranges are listed in Table \ref{S503}%
. It was found that the values of $\left\langle \Gamma \right\rangle $ are
consistent and the results of $\mu _{2}$ strongly depend on the delay time
range being fit. The fit values of the simulated data produced using the
mean static radius 260 nm and standard deviation 130 nm stabilize as the
delay time range is increased, but the relative width of apparent
hydrodynamic radius obtained using Eq. \ref{Rhwidth} is only about 0.23. The
fit results for $g^{\left( 2\right) }\left( \tau \right) $ are shown in Fig.
5. For both the fit results, the residuals are random.

\begin{table}[th]
\begin{center}
\begin{tabular}{|c|c|c|c|c|c|}
\hline
Delay time (s) & $\left\langle\Gamma\right\rangle _{first}\left(
s^{-1}\right) $ & $\chi ^{2}$ & $\left\langle \Gamma\right\rangle
_{two}\left( s^{-1}\right) $ & $\mu _{2}\left( s^{-2}\right) $ & $\chi ^{2}$
\\ \hline
2$\times$10$^{-7}$ to 0.004 & 216.9 $\pm$ 0.1 & 1.12 & 218.6 $\pm$ 0.5 & 
1009 $\pm$ 326 & 1.08 \\ \hline
2$\times$10$^{-7}$ to 0.00442 & 216.9 $\pm$ 0.1 & 1.07 & 218.4 $\pm$ 0.5 & 
868 $\pm$ 302 & 1.03 \\ \hline
2$\times$10$^{-7}$ to 0.00642 & 216.9 $\pm$ 0.1 & 1.02 & 218.0 $\pm$ 0.4 & 
598 $\pm$ 255 & 1.00 \\ \hline
2$\times$10$^{-7}$ to 0.00842 & 217.0 $\pm$ 0.1 & 0.98 & 217.0 $\pm$ 0.2 & 8 
$\pm$ 89 & 0.99 \\ \hline
2$\times$10$^{-7}$ to 0.01042 & 216.83 $\pm$ 0.04 & 1.20 & 217.1 $\pm$ 0.2 & 
59 $\pm$ 30 & 1.19 \\ \hline
2$\times$10$^{-7}$ to 0.01242 & 216.75 $\pm$ 0.04 & 1.25 & 217.3 $\pm$ 0.1 & 
97 $\pm$ 27 & 1.20 \\ \hline
2$\times$10$^{-7}$ to 0.01442 & 216.73 $\pm$ 0.04 & 1.23 & 217.3 $\pm$ 0.1 & 
106 $\pm$ 25 & 1.16 \\ \hline
\end{tabular}%
\end{center}
\caption{The fit results of simulated data produced based on the mean static
radius 50 nm and standard deviation 3 nm at different delay time ranges
using Eq. \protect\ref{cum} with $\protect\mu _{2}=0$ and $\protect\mu %
_{2}\neq 0$ respectively.}
\label{S503}
\end{table}

From the analysis of the simulated data, the width of apparent hydrodynamic
radius cannot be determined for narrow distributions because the values of $%
\mu _{2}$ have a strong dependence on the noises, errors or delay time range
being fit and can be determined for wide particle size distributions since
the values of $\left\langle \Gamma \right\rangle $ and $\mu _{2}$ stabilize
as the delay time range being fit is increased. However, the value of the
relative width of apparent hydrodynamic radius obtained using the DLS
technique has largely different from the relative width of the simulated
data being fit. Because the standard method of cumulants obtains the
distribution of apparent hydrodynamic radius from the deviation between the
exponentiality and $g^{\left( 2\right) }\left( \tau \right) $ measured at a
single scattering angle, the simulated data were used to explore the
nonexponentiality of $g^{\left( 2\right) }\left( \tau \right) $ at a
scattering angle of 30$^{\text{o}}$. The logarithm of the simulated data
produced without noises and errors were plotted as a function of the delay
time. All results for the standard deviations 3 nm, 10 nm, 20 nm, 25 nm and
mean static radius 50 nm are shown in Fig. 6a and the standard deviations 16
nm, 52 nm, 104 nm, 130 nm and mean static radius 260 nm are shown in Fig.
6b, respectively. Figure 6 shows the effects of the standard deviation are
small on the nonexponentiality of $g^{\left( 2\right) }\left( \tau \right) $
and large on the initial slope of the logarithm of $g^{\left( 2\right)
}\left( \tau \right) $ $\left\langle \Gamma \right\rangle $ at a scattering
angle of 30$^{\text{o}}$.

Because the apparent hydrodynamic radius is related to $\left\langle \Gamma
\right\rangle $, the effects of the particle size distribution on the
apparent hydrodynamic radius were thus explored. The values of the apparent
hydrodynamic radius obtained using Eqs. \ref{cum} and \ref{Rhapp} with $\mu
_{2}=0$ and $\mu _{2}\neq 0$, and Eq. \ref{Rhallcal} respectively for the
simulated data produced using the mean static radius 50 nm and different
standard deviations are shown in Table \ref{distribution}. From the
relationship $a=R_{h}/R_{s}=1.1$, the mean hydrodynamic radius $\left\langle
R_{h}\right\rangle $ is 55 nm. The relative deviations $\left(
R_{h,app1}-\left\langle R_{h}\right\rangle \right) /\left\langle
R_{h}\right\rangle $ as a function of the standard deviation are also listed
in Table \ref{distribution}.

\begin{table}[th]
\begin{center}
\begin{tabular}{|c|c|c|c|c|}
\hline
$\sigma /\left\langle R_{s}\right\rangle $ & $R_{h,app1}\left( nm\right) $ & 
$R_{h,app2}\left( nm\right) $ & $R_{cal}\left( nm\right) $ & $\frac{%
R_{h,app1}-\left\langle R_h\right\rangle}{\left\langle R_h\right\rangle}$ \\ 
\hline
0.06 & 56.2 $\pm $ 0.2 & 56.2 $\pm $ 0.5 & 56.1 & 0.02 \\ \hline
0.1 & 57.9 $\pm $ 0.2 & 57.5 $\pm $ 0.2 & 57.9 & 0.05 \\ \hline
0.2 & 65.3 $\pm $ 0.1 & 64.6 $\pm $ 0.1 & 65.4 & 0.19 \\ \hline
0.3 & 75.5 $\pm $ 0.5 & 73.8 $\pm $ 0.4 & 75.5 & 0.37 \\ \hline
0.4 & 87.1 $\pm $ 0.6 & 84.4 $\pm $ 0.5 & 86.8 & 0.58 \\ \hline
0.5 & 99.8 $\pm $ 0.8 & 96.3 $\pm $ 0.8 & 98.7 & 0.81 \\ \hline
0.6 & 112. $\pm $ 1. & 108.1 $\pm $ 0.9 & 110.9 & 1.04 \\ \hline
0.7 & 125.0 $\pm $ 0.7 & 120. $\pm $ 1. & 123.2 & 1.27 \\ \hline
\end{tabular}%
\end{center}
\caption{Values of $R_{h,app}$ for the simulated data produced using the
mean static radius 50 nm and different standard deviations, and relative
deviations $\left( R_{h,app1}-\left\langle R_{h}\right\rangle \right)
/\left\langle R_{h}\right\rangle $.}
\label{distribution}
\end{table}

The results in Table \ref{distribution} show that the value of the apparent
hydrodynamic radius is greatly influenced by the particle size distribution.
The part of apparent hydrodynamic radius represents the effects of particle
size distributions. The wider the particle size distribution, the larger the
value of the apparent hydrodynamic radius. The consistency between the value
calculated from Eq. \ref{Rhallcal} and the result obtained using the first
cumulant also shows the deviation between the exponentiality and $g^{\left(
2\right) }\left( \tau \right) $ at a scattering angle of 30$^{\text{o}}$ is
small even for very wide distribution like the relative width distribution
50\%. The difference between the results obtained using the first and first
two cumulants is influenced by the particle size distribution. For narrow
distributions, they are almost equal. For a wide distribution like 50\%, the
difference is less than 4\%. The relative width of apparent hydrodynamic
radius obtained from the deviation between the exponentiality and $g^{\left(
2\right) }\left( \tau \right) $ is about 24\% for the simulated data
produced using the relative width of hydrodynamic radius 50\%. This
difference is caused by the method of cumulants. The average decay rate of $%
g^{\left( 2\right) }\left( \tau \right) $ is determined by the particle size
distribution and the relationship between the static and hydrodynamic radii
together and the method of cumulants measures the width of apparent
hydrodynamic radius from the nonexponentiality of $g^{\left( 2\right)
}\left( \tau \right) $ related to the exponentiality of the average decay
rate at a single scattering angle.

\section{CONCLUSION}

The effects of the particle size distribution on the nonexponentiality of $%
g^{\left( 2\right) }\left( \tau \right) $ are small and very large on the
initial slope of the logarithm of $g^{\left( 2\right) }\left( \tau \right) $
at a scattering angle of 30$^{\text{o}}$. The values of the apparent
hydrodynamic radius are also greatly influenced by the particle size
distribution. The wider the particle size distribution, the larger the value
of the apparent hydrodynamic radius. The questions that the standard DLS
techniques are not suited to the accurate determination for the
poly-dispersities (standard deviation/mean size) less than 10\% and the
distribution of apparent hydrodynamic radius is different from the
distribution of hydrodynamic radius of particles are caused by the method of
cumulants. The average decay rate (the initial slope of the logarithm) of $%
g^{\left( 2\right) }\left( \tau \right) $ is determined by the particle size
distribution and the relationship between the static and hydrodynamic radii
together and the method of cumulants detects the width of apparent
hydrodynamic radius from the nonexponentiality of $g^{\left( 2\right)
}\left( \tau \right) $ related to the exponentiality of the average decay
rate at a single scattering angle. The situations that the standard method
of cumulants are used at other scattering angles will be studied further.

Fig .1. The experimental data and fit results of $g^{\left( 2\right) }\left(
\tau \right) $ for Latex-1 at a scattering angle of 30$^{\text{o}}$ and a
temperature of 298.45 K. The circles show the experimental data, the line
represents the fit results obtained using Eq. \ref{cum} and the diamonds
show the residuals $\left( y_{i}-y_{fit}\right) /\sigma _{i}$. The results
for $\mu _{2}=0$ and $\mu _{2}\neq 0$ are shown in a and b respectively.

Fig. 2. The experimental data and fit results of $g^{\left( 2\right) }\left(
\tau \right) $ for Latex-1 at a scattering angle of 90$^{\text{o}}$ and a
temperature of 298.45 K. The circles show the experimental data, the line
represents the fit results obtained using Eq. \ref{cum} and the diamonds
show the residuals $\left( y_{i}-y_{fit}\right) /\sigma _{i}$. The results
for $\mu _{2}=0$ and $\mu _{2}\neq 0$ are shown in a and b respectively.

Fig. 3. The experimental data and fit results of $g^{\left( 2\right) }\left(
\tau \right) $ for PNIPAM-1 at a scattering angle of 30$^{\text{o}}$ and a
temperature of 302.38K. The circles show the experimental data, the line
represents the fit results obtained using Eq. \ref{cum} and the diamonds
show the residuals $\left( y_{i}-y_{fit}\right) /\sigma _{i}$. The results
for $\mu _{2}=0$ and $\mu _{2}\neq 0$ are shown in a and b respectively.

Fig. 4. The fit results of $g^{\left( 2\right) }\left( \tau \right) $ for
the simulated data produced based on the mean static radius 50 nm and
standard deviations 10 nm at a scattering angle of 30$^{\text{o}}$. The
circles show the simulated data, the line represents the fit results
obtained using Eq. \ref{cum} and the diamonds show the residuals $\left(
y_{i}-y_{fit}\right) /\sigma _{i}$. The results for $\mu _{2}=0$ and $\mu
_{2}\neq 0$ are shown in a and b respectively.

Fig. 5. The fit results of $g^{\left( 2\right) }\left( \tau \right) $ for
the simulated data produced with the mean static radius 260 nm and standard
deviations 130 nm at a scattering angle of 30$^{\text{o}}$. The circles show
the simulated data, the line represents the fit results obtained using Eq. %
\ref{cum} and the diamonds show the residuals $\left( y_{i}-y_{fit}\right)
/\sigma _{i}$. The results for $\mu _{2}=0$ and $\mu _{2}\neq 0$ are shown
in a and b respectively.

Fig. 6. The differences between the lines and plots of $\ln\left( \left(
g^{\left( 2\right) }\left( \tau \right) -1\right) /\beta \right) $ as a
function of the delay time. The symbols show the simulated data and the
lines show the linear fitting to the simulated data respectively. The
results for the simulated data produced using the mean static radius 50 nm
and 260 nm are shown in a and b respectively.

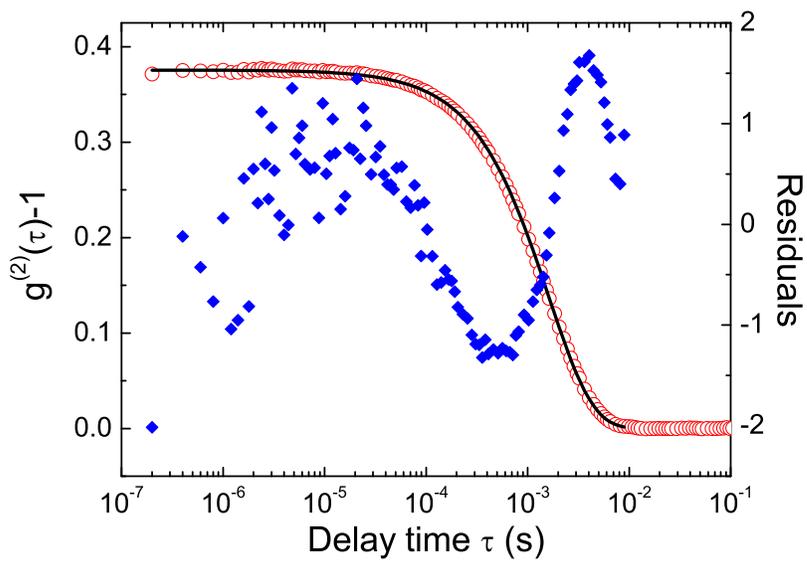
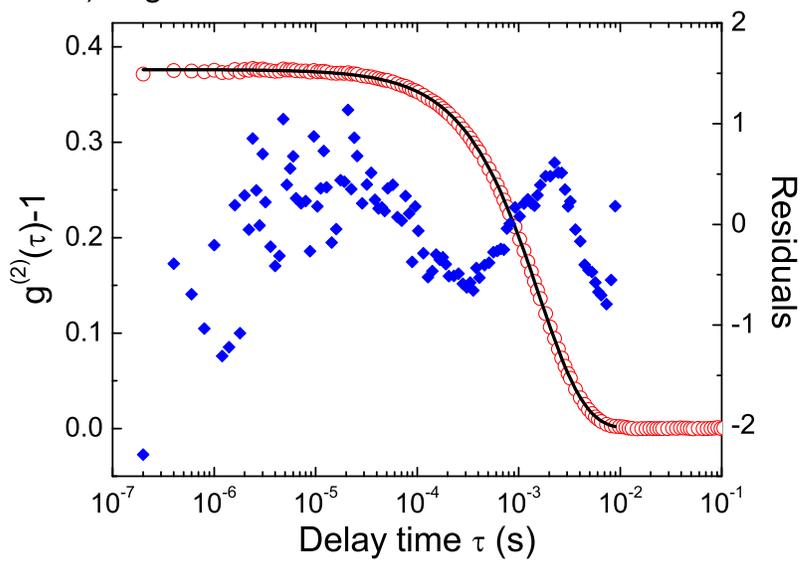
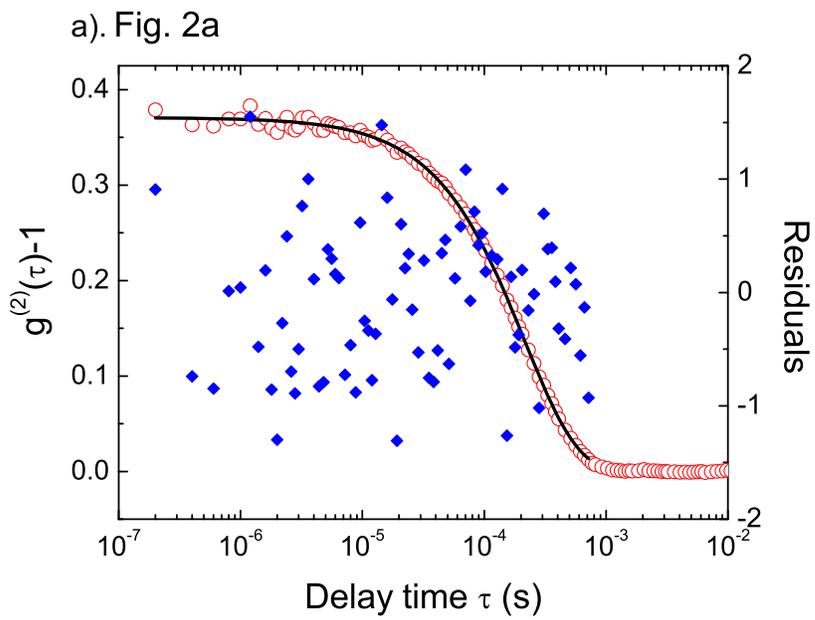
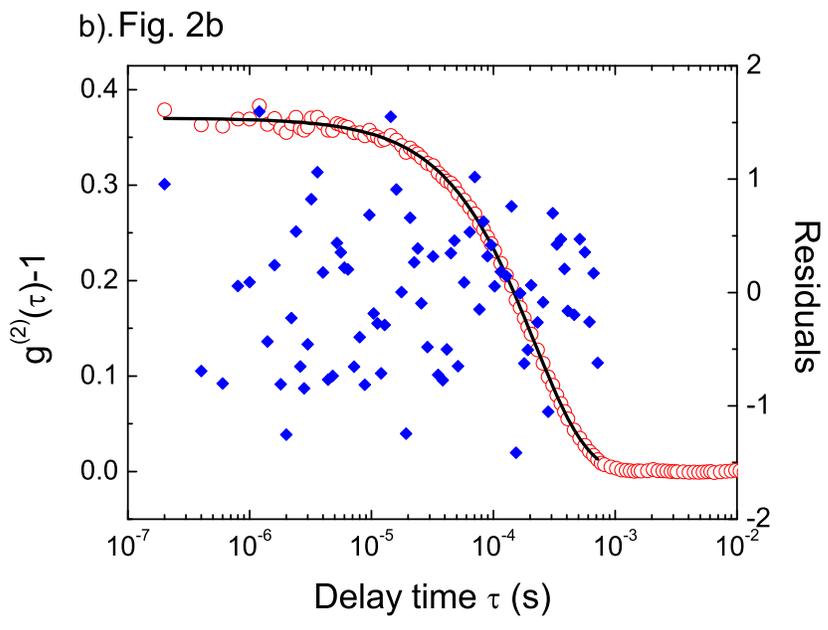
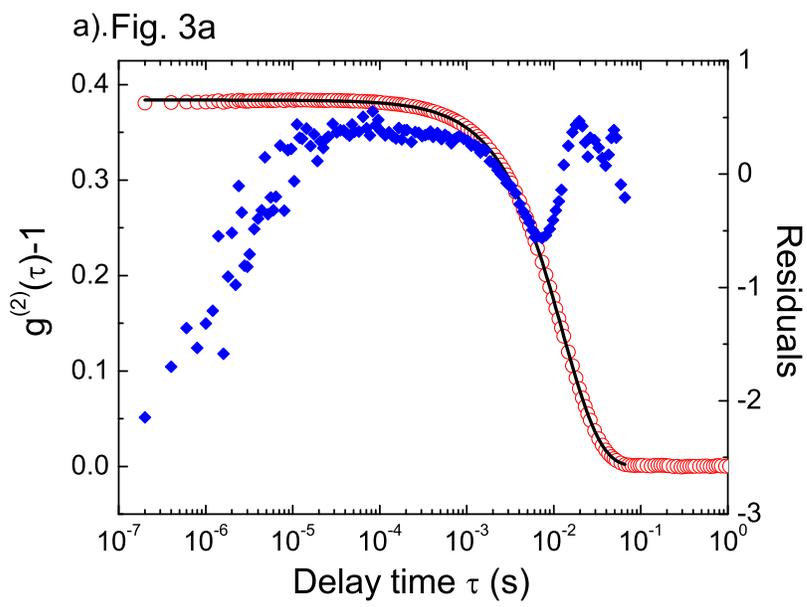
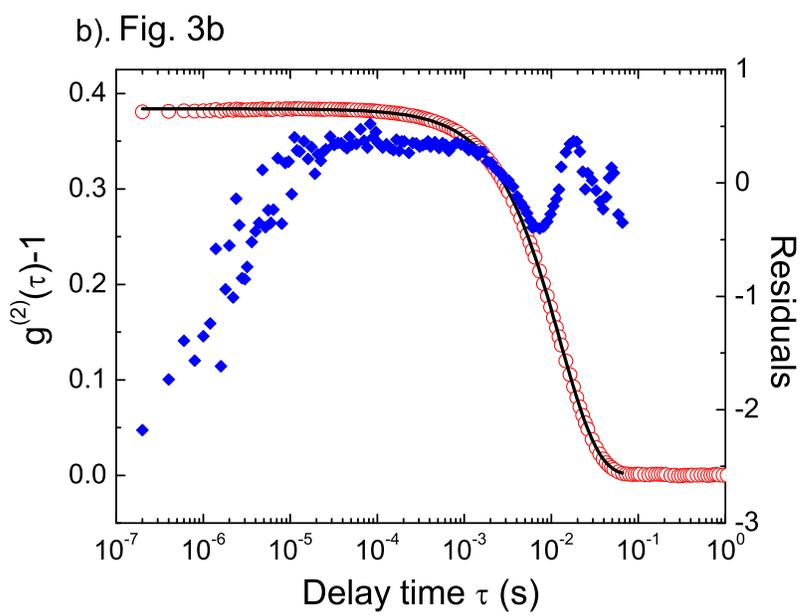
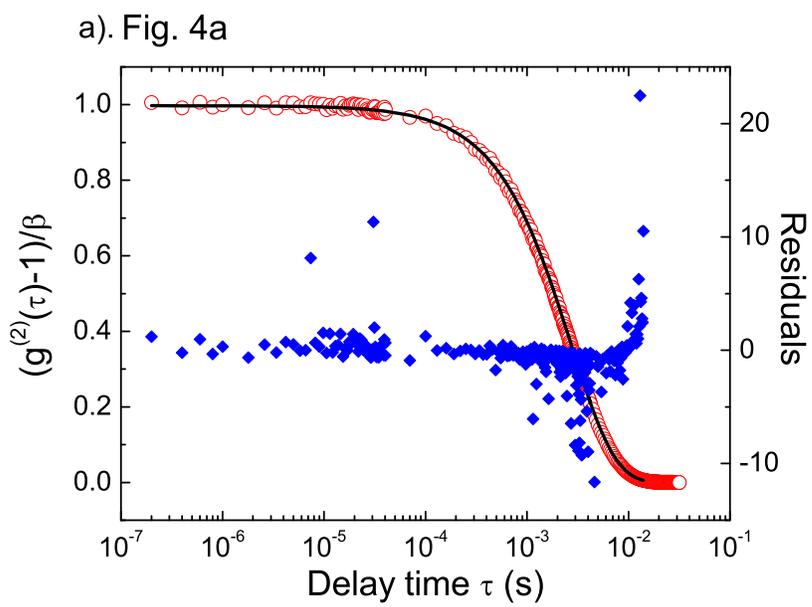
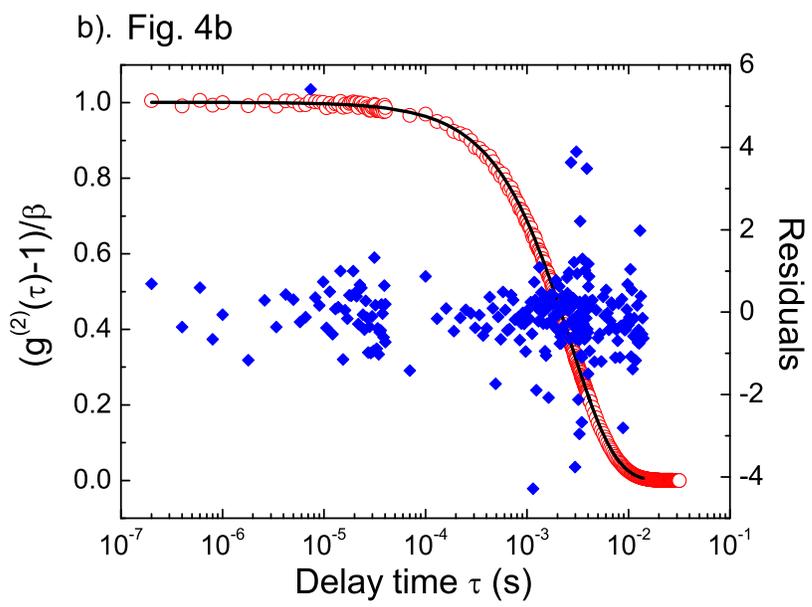
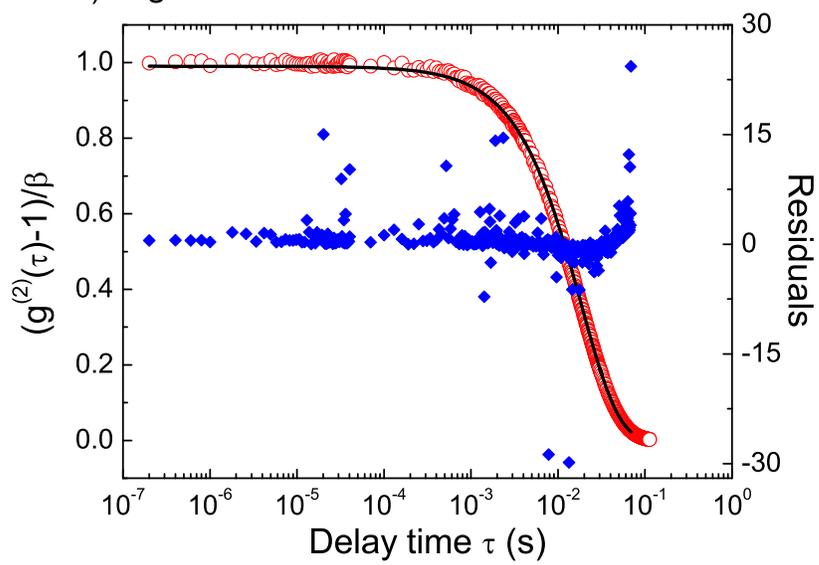
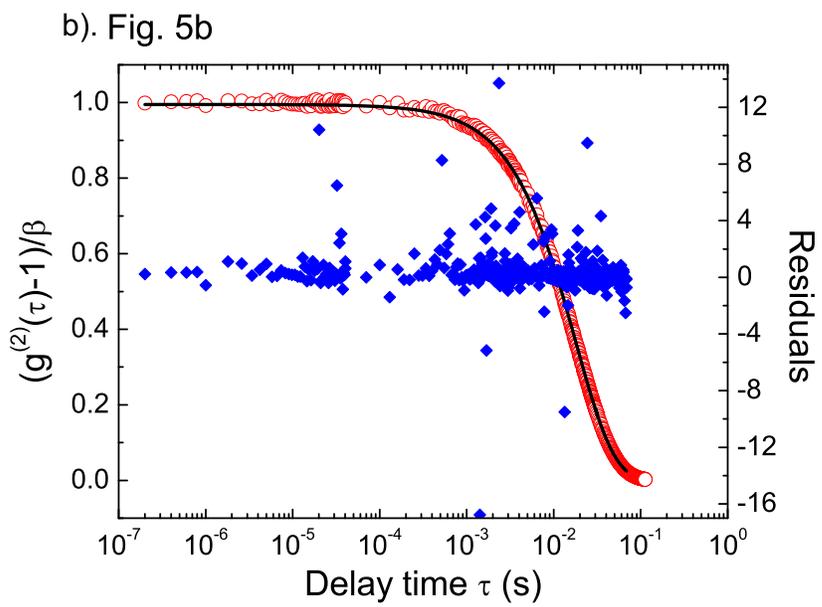
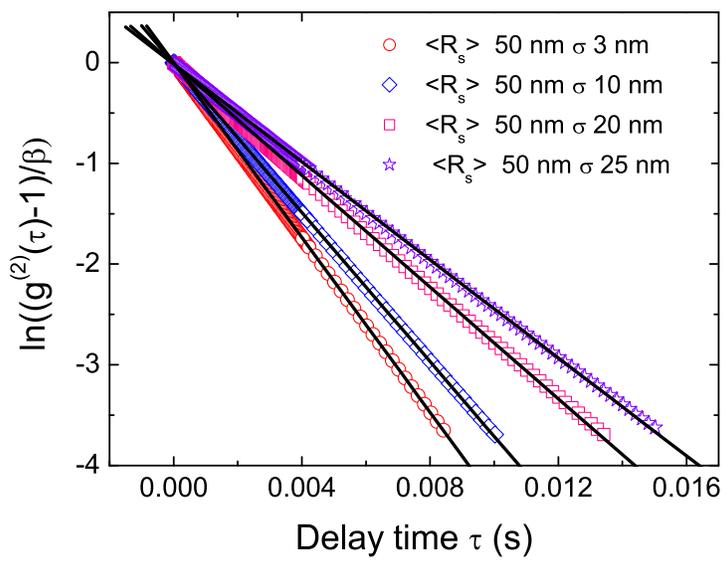
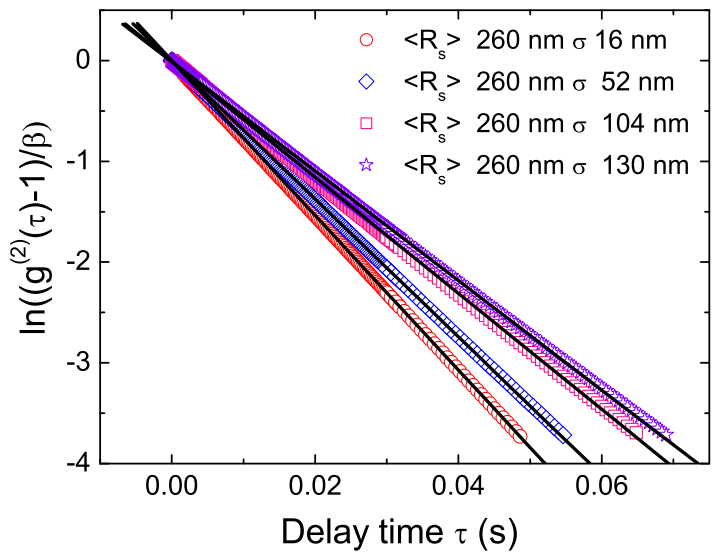
\end{document}